\documentclass[a4paper,aps,prb,amsmath,reprint,superscriptaddress]{revtex4-1}
\usepackage[latin9]{inputenc}
\usepackage{textcomp}
\usepackage{amsbsy}
\usepackage{graphicx}
\usepackage{subscript}

\makeatletter

\DeclareFontEncoding{LGR}{}{}
\DeclareRobustCommand{\greektext}{%
  \fontencoding{LGR}\selectfont\def\encodingdefault{LGR}}
\DeclareRobustCommand{\textgreek}[1]{\leavevmode{\greektext #1}}
\ProvideTextCommand{\~}{LGR}[1]{\char126#1}

\DeclareFontEncoding{LGR}{}{}
\DeclareRobustCommand{\greektext}{%
  \fontencoding{LGR}\selectfont\def\encodingdefault{LGR}}
\DeclareRobustCommand{\textgreek}[1]{\leavevmode{\greektext #1}}
\ProvideTextCommand{\~}{LGR}[1]{\char126#1}

\usepackage{dcolumn}
\usepackage{bm}
\usepackage[mathlines]{lineno}

\makeatother

\begin{document}

\title{Optical probing of Rayleigh wave driven magneto-acoustic resonance}

\author{P. Kuszewski}

\affiliation{Sorbonne Université, CNRS, Institut des Nanosciences de Paris, 4
place Jussieu,75252 Paris France}

\author{J.-Y. Duquesne}

\affiliation{Sorbonne Université, CNRS, Institut des Nanosciences de Paris, 4
place Jussieu,75252 Paris France}

\author{L. Becerra}

\affiliation{Sorbonne Université, CNRS, Institut des Nanosciences de Paris, 4
place Jussieu,75252 Paris France}

\author{A. Lemaître}

\affiliation{Centre de Nanosciences et de Nanotechnologies, CNRS, Univ. Paris-Sud,
Université Paris-Saclay, 91460 Marcoussis, France}

\author{S. Vincent}

\affiliation{Sorbonne Université, CNRS, Institut des Nanosciences de Paris, 4
place Jussieu,75252 Paris France}

\author{S. Majrab}

\affiliation{Sorbonne Université, CNRS, Institut des Nanosciences de Paris, 4
place Jussieu,75252 Paris France}

\author{F. Margaillan}

\affiliation{Sorbonne Université, CNRS, Institut des Nanosciences de Paris, 4
place Jussieu,75252 Paris France}

\author{C. Gourdon}

\affiliation{Sorbonne Université, CNRS, Institut des Nanosciences de Paris, 4
place Jussieu,75252 Paris France}

\author{L. Thevenard}

\affiliation{Sorbonne Université, CNRS, Institut des Nanosciences de Paris, 4
place Jussieu,75252 Paris France}

\date{\today}
\begin{abstract}
The resonant interaction of electrically excited travelling surface
acoustic waves and magnetization has been hitherto probed through
the acoustic component. In this work it is investigated using time-resolved
magneto-optical detection of magnetization dynamics. To that end,
we develop an experimental scheme where laser pulses are used both
to generate the acoustic wave frequency and to probe magnetization
dynamics thus ensuring perfect phase locking. The light polarization
dependence of the signal enables to disentangle elasto-optic and magneto-optic
contributions and to obtain the in-plane and out-of-plane dynamic
magnetization components. Magnetization precession is proved to be
driven solely by the acoustic wave. Its amplitude is shown to resonate
at the same field at which we detect piezo-electrically the resonant
attenuation of the acoustic wave, clearly evidencing the magneto-acoustic
resonance with high sensitivity. 
\end{abstract}

\date{\today}

\maketitle

\section{Introduction}

The recent years have witnessed a renewed and growing interest in
the use of acoustic waves to excite spin waves as an alternative,
fast, efficient, and heat-free means to generate and control (possibly
coherently and remotely) information in magnetic materials \cite{Dreher2012,Bombeck2012,Li2014b,Thevenard2014,Yahagi2014,Shen2015a,Davis2015,Gowtham2015,Chowdhury2015,Janusonis2015a,Thevenard2016a,Thevenard2016,Chang2017a,Hashimoto2017a,Kim2017,Kuszewski2018}.
Magnetization switching \cite{Davis2015,Li2014b,Thevenard2016a,Thevenard2016,Kuszewski2018},
and parametric excitation of spin waves \cite{Chang2017a,Chowdhury2015}
have for instance been demonstrated, thus opening new perpectives
for applications. The efficiency of acoustic waves relies on magneto-elastic
coupling that occurs in magnetostrictive materials. Static strain
contributes to magnetic anisotropy \cite{Lemaitre2008,Nath1999} whereas
dynamic strain (acoustic waves) exerts a torque on the magnetization
leading to precessional motion predicted to be resonantly enhanced
when the spin wave frequency matches the acoustic wave frequency \cite{Linnik2011,Thevenard2013a}.

Among the various acoustic waves, propagating surface acoustic waves
(SAW), such as Rayleigh waves, with typical frequencies up to the
GHz range and dynamical strain component confined within a few \textmu m
from the surface, are particularly well suited to excite magnetization
dynamics in thin ferromagnetic layers on a substrate. SAWs are easy
to implement thus finding applications in various fields of soft and
condensed matter physics \cite{Yeo2014,Mamishev2004,Poole2017,DeLima2005}.
They can be generated either optically via thermoelasticity \cite{Neubrand1992,Yahagi2014,Janusonis2015a,Chang2017a}
or electrically via the piezoelectric effect using interdigitated
transducers (IDTs) \cite{Royer2}. Since electrically generated SAWs
are a well-mastered technology in microelectronics, SAW-induced magnetization
control could be readily implemented in logic or memory devices, with
the unprecedented possibility to use wave physics tools such as focusing,
diffraction, and wave-guiding to address a magnetic bit.

In ferromagnetic systems, interaction of electrically generated SAWs
with magnetization was evidenced by piezo-electrical detection after
propagation of the SAW on the sample surface \cite{Dreher2012,Thevenard2014}.
The SAW amplitude and velocity resonantly decreased when the frequency
of magnetization precession was varied through the SAW frequency by
an applied magnetic field. By analogy with cavity-ferromagnetic resonance
(FMR) this is known as SAW-FMR. However, magnetization precession
has not been concomitantly observed although time-resolved detection
of magneto-optical effects using probe laser pulses could provide
a direct and most sensitive method to access magnetization dynamics
\cite{Shihab2017}. This requires a non-trivial synchronization of
the SAW generation with probe laser pulses. This difficulty has been
up to now bypassed by utilizing optically generated stationary SAWs
\cite{Janusonis2015a,Yahagi2014}. Very recently, synchrotron X-ray
probe pulses were locked to electrically generated SAWs to investigate
strain-driven modes in patterned ferromagnetic samples \cite{Foerster2017}.
No (resonant) magnetization dynamics were observed however.

In this article, we demonstrate the forced magnetic precession induced
by \textit{travelling} SAW bursts, using a table-top optical set-up.
To that end we implement an innovative and simple way to generate
electrically SAW bursts by IDTs with perfect phase locking to laser
pulses, which allows the optical investigation of time- and space-resolved
magnetization dynamics excited by the SAW. We illustrate the potentiality
of this technique with a ferromagnetic semiconductor layer of GaMnAs
on piezoelectric GaAs, taking advantage of its low FMR frequency easily
tunable across the SAW frequency, and sizable magneto-optical effects.
Using the probe light polarization and the magnetic field dependencies
of the time-resolved signal we separate the magneto-optical from the
pure elasto\textendash optical contribution. The in-plane and out-of
plane magnetization dynamical components display the predicted dependencies
on the magnetic field. Finally, we show that resonant magnetic excitation
appears concomitantly with resonant SAW absorption, the detection
of the former being more sensitive than the latter.\\

\section{Experimental setup}

The simplest IDT consists of two comb-shaped electrodes in zipper
configuration, that work as capacitors on a piezoelectric surface.
When a radio-frequency (RF) voltage at $f_{SAW}$ is applied, SAWs
are generated and propagate on either side of the IDT. The wavelength
$\lambda_{SAW}$ of the excited SAW is determined by the period of
the electrode teeth. Frequency and wavelengeth are related by $\lambda_{SAW}=\frac{v_{R}}{f_{SAW}}$
where $v_{R}$ is the Rayleigh velocity. Typical $\lambda_{SAW}$
values are around a few microns. The strain components excited with
the IDT are: longitudinal $\varepsilon_{xx}\:(x\parallel\mathbf{k}_{SAW})$,
transverse $\varepsilon_{zz}$ (\textit{z} perpendicular to the surface)
and shear $\varepsilon_{xz}$ (negligibly small close to the surface)
\cite{Royer2}. In more sophisticated IDT architectures it is possible
to excite overtone harmonics of the fundamental frequency, such as
the split-52 design \cite{Schulein2015}, which is used in this work.
\begin{figure*}[htbp]
\includegraphics[scale=0.4]{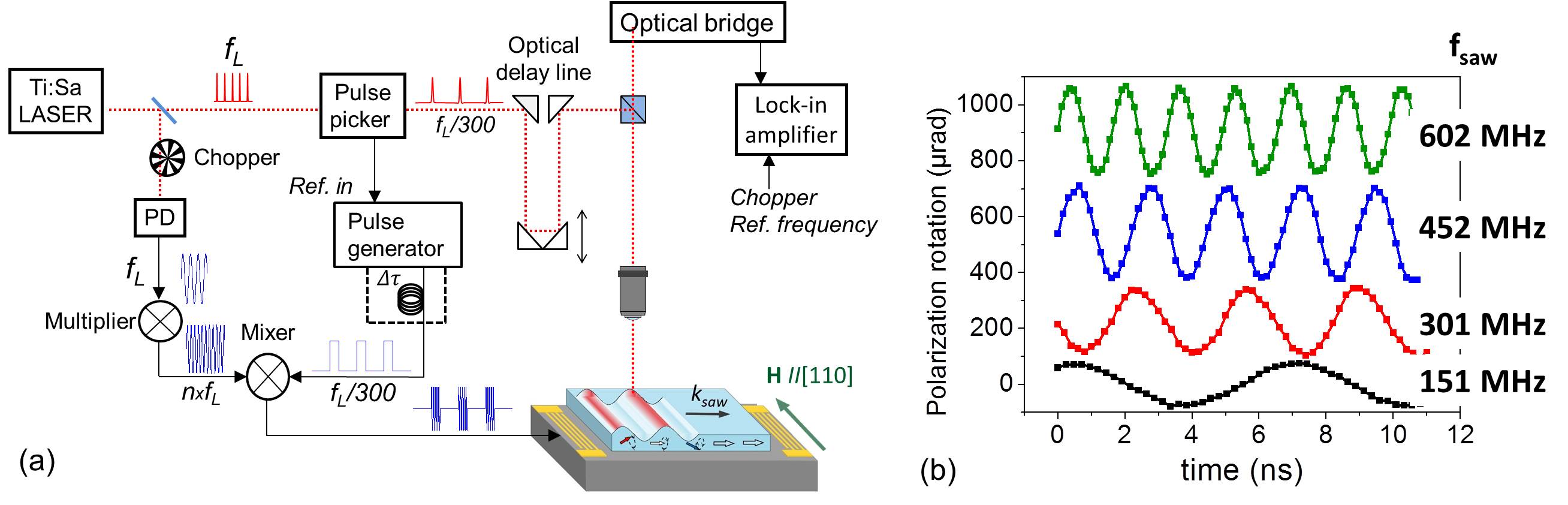}\caption{\label{fig:sch=00003D0000E9ma}(a) Schematics of the experimental
set-up, (b) Time-resolved polarization rotation signal for four SAW
frequencies (vertically off-set for clarity). The probe polarization
was set at 60\textdegree{} with respect to $\mathbf{k}_{SAW}$. The
applied magnetic field was 19.5 mT. The temperature was T=60K.}
\end{figure*}
Optical time-resolved measurement of magnetization dynamics requires
a fixed phase between the laser pulses and the acoustic bursts. The
most common solution is a phase-locked loop \cite{Hollander2017}.
The `master' clock imposes the repetition rate of the `slave' laser.
The laser cavity length is continuously adjusted in order to meet
the set frequency. We propose a simpler, elegant, and low-cost alternative
approach where $f_{SAW}$ is built from the laser repetition rate
thereby ensuring a stable phase lock. Two research groups have very
recently reported independently similar synchronization methods \cite{Foerster2017,Weiss2018}.

In order to generate the SAW RF frequency, a small fraction of a Ti:Sapphire
laser at $\sim$75 MHz repetition rate ($f_{L}$) was directed to
a Silicon photodiode (Thorlabs PDA10A-EC, bandwidth 150 MHz) to pick
up the laser fundamental frequency (Fig. 1(a)). Using different RF
multipliers, multiples of the laser frequency were then generated:
$n\thinspace f_{L}$, where $n=2,\thinspace4,\thinspace6,\,8$. To
ensure that $n\thinspace f_{L}$ is free of any harmonics, the signal
was filtered with 50 dB rejection band pass filters. The signal level
was adjusted with a linear 40 dB amplifier and a set of attenuators.

The larger fraction of the 130 fs laser pulses (wavelength 722 nm)
was sent to a pulse selector (Fig. 1(a)). The repetition rate was
reduced by a factor 300 to $f_{rep}=$ 250 kHz. This beam was used
to probe the magnetization and strain dynamics in a time window of
12 ns controlled with a motorized optical delay line stage.

As we shall see hereafter the IDT needs to be fed with a pulsed RF
signal. In order to fabricate these pulses the $n\thinspace f_{L}$
CW RF signal was mixed with a 400 ns-wide rectangular pulse train
from a pulse generator (Keysight 81150A) at the pulse picker frequency
$f_{rep}$ thus ensuring that both the envelope frequency and the
RF frequency of the SAW are commensurate with $f_{L}$. With the pulse
generator we could modify the burst duration and its arrival time
(electronic delay) to the transducer. The `edge-to-reference` jitter
\cite{maichen2006} defined as the timing variation between the rising
edge of the rectangular pulse and the optical pulse shows a standard
deviation $\sigma=13.1$ ps comfortably below 1\% of the highest SAW
period.

The linearly polarized laser pulses at $f_{rep}$ were focused on
the sample surface with a long working distance objective lens with
numerical aperture 0.4. The laser beam reflected off the sample traveled
through the same objective. The polarization rotation induced by the
dynamical magnetization and strain components was detected by a balanced
optical bridge relying on magneto-optical effects\cite{Shihab2017,Nemec2012}
and the photo-elastic (PE) effect \cite{Royer2}, respectively. The
bridge output was demodulated by a lock-in amplifier at the frequency
of a mechanical chopper $f_{ch}$=541 Hz inserted on the laser beam
before the slow photodiode (Fig. 1(a)),\textit{ i.e.} modulating the
excitation SAW. Polarization rotations as small as 0.1 \textmu rad
can routinely be measured.

The sample was a 45 nm thick, in-plane magnetized Ga\textsubscript{0.95}Mn\textsubscript{0.05}As
layer on a (001) GaAs substrate, annealed for 16 h at $200^{o}$C,
with a Curie temperature $T_{C}=120$ K. It was placed in an Oxford
MicroStat HiRes cryostat that ensures the required mechanical stability
and optical access. The GaMnAs layer has a strong uniaxial magnetic
anisotropy with the easy axis along {[}1-10{]} and exhibits a large
magneto-elastic coupling \cite{Kuszewski2018}. An in-plane magnetic
field aligned with the hard axis {[}110{]} was applied to decrease
the precession frequency and make it cross the SAW frequency. A $\mathrm{2x2\:mm}^{2}$
square mesa of GaMnAs was prepared by chemical etching. At two opposite
ends a set of 42 nm thick Aluminum IDTs with 1 mm aperture were deposited
on the GaAs (Fig. \ref{fig:sch=00003D0000E9ma}(a)). In order to provide
efficient operation of the IDTs even when varying the temperature
(given the temperature dependence of $v_{R}(T)$) and to achieve a
compromise between excitation efficiency and bandwidth, the IDTs were
designed as 25 pairs of split-52 electrodes. With a digit width and
inter-digit spacing of 1.9 \textmu m, the fundamental period was $\lambda_{SAW}$=19
\textmu m. This ensured that the wavelength of all excited SAW frequencies
was substantially larger than the laser spot diameter (full-width
at half maximum of about 1 \textmu m). The RF voltage applied to the
IDT triggers the propagation of a SAW along the {[}1-10{]} axis. The
RF power was set to 25 dBm for most data presented here. We used the
second IDT to detect the SAW electrically by the inverse piezoelectric
effect after it had travelled $d=2$ mm along the layer surface. Figure
\ref{fig:burst} presents an oscilloscope trace of the recorded signal
(red line). The first 400 ns burst is the electromagnetic radiation.
It travels with the speed of light and appears immediately at the
receiver. After around $t_{trans}=\frac{d}{v_{R}}=760$ ns the acoustic
echo arrives. The pulsed RF allows to separate in time the electromagnetic
radiation and the acoustic echo. The risetime of the acoustic echo
is defined by the IDT geometry, the SAW velocity, and the burst duration
\cite{Royer2}.

\section{Results and discussion}

Figure \ref{fig:sch=00003D0000E9ma}(b) shows the time-resolved polarization
rotation (TRPR) at 60 K for excitation SAW frequencies 151, 301, 452
and 602 MHz as a function of the optical delay. The frequency of the
detected signals matches exactly the excitation frequency. The amplitude
of the signal varies with the magnetic field up to $T_{C}$, above
which we observe only a field independent signal. This proves that
the TRPR signal contains both a magnetic (field-dependent) contribution
and a non-magnetic one arising from the elasto-optic effect.

\begin{figure}
\includegraphics[scale=0.58]{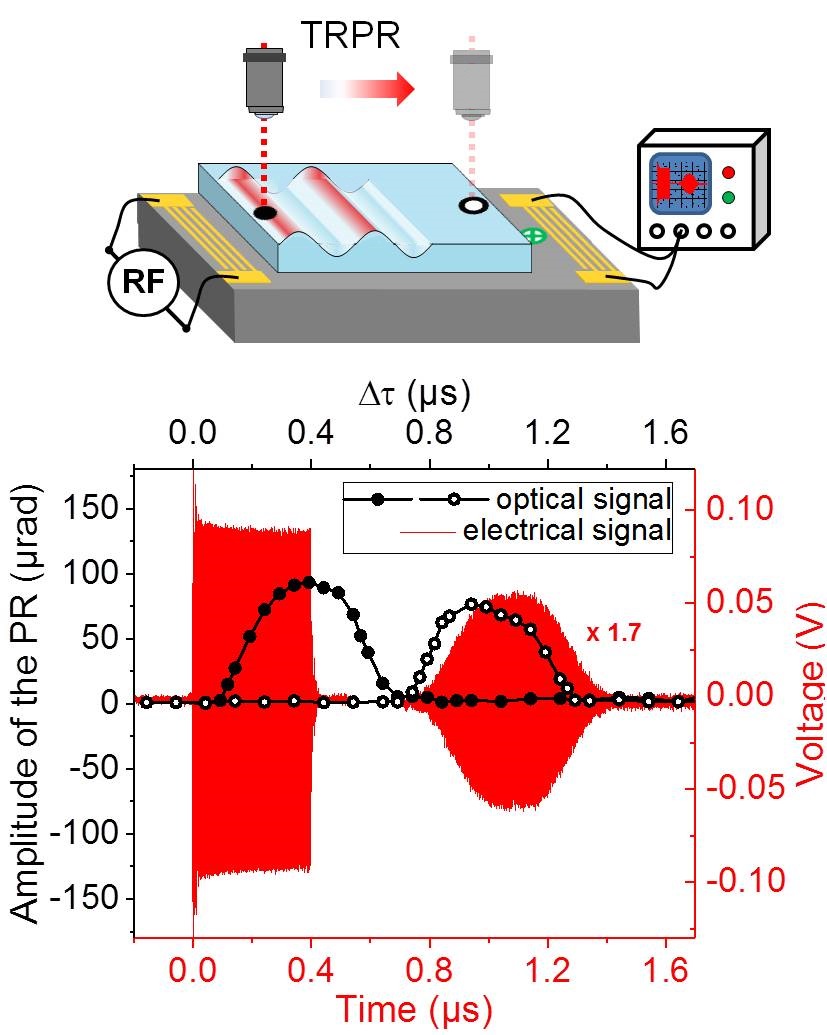}\caption{\label{fig:burst}Time-domain signal detected electrically (red solid
line) by an oscilloscope, and amplitude of the optically detected
TRPR oscillations (circles) as a function of the electronic delay.
TRPR was measured in two places on the sample (open and closed symbols
on the scheme). The crossed-circle symbol indicates where the optical
SAW-FMR data shown in Fig. 4(h) was taken. The surface amplitudes
of the excited SAW strain components are $\varepsilon_{xx}=4.8\times10^{-5}$
and $\varepsilon_{zz}=2.2\times10^{-5}$ estimated following the procedure
described in ref. \cite{Kuszewski2018}. The SAW frequency is 452
MHz, the applied field is $\mu_{0}H=$20 mT, the polarization angle
is $\beta$=30 deg. }
\end{figure}

In order to prove that the optically detected dynamics is indeed triggered
by the SAW we monitored the amplitude of the TRPR signal at various
delays after the beginning of the the RF burst by setting an electronic
delay $\Delta\tau$ between the electric excitation and the optical
pulse probe (Fig. 1(a)). First the probe beam was set at 100 \textmu m
from the receiving IDT (1.9 mm from the emitter, open circle in the
sample scheme of Fig. \ref{fig:burst}). The open circle symbol curve
represents the amplitude of the TRPR oscillations versus the electronic
delay. It has the same shape as the acoustic echo and peaks slightly
before the arrival of this echo on the receiving IDT, in agreement
with the location of the detection spot. Then the probe beam was positioned
67 \textmu m from the emitter IDT (full circle in the sample scheme
of Fig. \ref{fig:burst}). The full circle curve represents the amplitude
of the oscillations versus the electronic delay. It also has the same
shape as the acoustic echo and is detected earlier, in agreement with
the position of the detection spot. The 20$\%$ burst amplitude decrease
from one burst to the other can be directly correlated to the loss
of acoustic amplitude via magneto-elastic interaction, as will be
seen further (Fig. 4(g)). For both spot positions, no signal is observed
at the arrival time of the electromagnetic radiation. These results
show that the TRPR signal is indeed generated by the SAW.

In order to disentangle the magneto-optic and the elasto-optic contributions
in the TRPR signal we analyze the dependence of the signal on the
probe light polarization and the applied magnetic field. The elasto-optic
effect gives rise to a dynamic birefringence with axes parallel and
perpendicular to the SAW wavevector $\boldsymbol{k}_{SAW}$. The resulting
rotation of linear polarization is proportional to the $P_{44}$ component
of the elasto-optic tensor of the cubic GaAs and the $\varepsilon_{xx}$
strain component \cite{Royer2000a,Saito2010,Santos1999}. It does
not depend on the magnetic field. Besides we expect the contribution
from two magneto-optical effects \cite{Nemec2012,Shihab2017} : the
polar magneto-optical Kerr effect (PMOKE) sensitive to the out-of-plane
dynamic component of the magnetization $\delta\theta$ and independent
of the incoming polarization and the (weaker) Voigt effect (magnetic
linear dichroism (MLD)) sensitive to the in-plane dynamic component
of the magnetization $\delta\phi$ and to the field-dependent in-plane
equilibrium orientation of the magnetization $\phi_{0}(H)$. The TRPR
signal can therefore be expressed as \cite{Shihab2017,Nemec2012,Royer2}:
\begin{eqnarray}
\delta\beta & = & K\delta\theta(H,t)+2V\delta\phi(H,t)\thinspace\cos\left(2\left(\beta-\phi_{0}(H)\right)\right)\nonumber \\
 &  & +P_{E}\varepsilon_{xx}(t)\thinspace\sin2\beta\\
 & = & K\delta\theta(H,t)+2V\delta\phi(H,t)\cos2\phi_{0}(H)\thinspace\cos2\beta\nonumber \\
 &  & +\left(2V\delta\phi(H,t)\,\sin2\phi_{0}(H)+P_{E}\varepsilon_{xx}(t)\right)\thinspace\sin2\beta
\end{eqnarray}
where \textgreek{b} is the angle of the probe polarization with respect
to $\boldsymbol{k}_{SAW}$, $K$ and $V$ are the Kerr and Voigt magneto-optical
coefficients, respectively, and $P_{E}=Re\left(n^{3}P_{44}/(n^{2}-1)\right)$
is the elasto-optic coefficient with $n$ the refractive index. Indeed
the experimental TRPR signal shows a clear dependence on the probe
polarization (Fig. 3(a)) and on the magnetic field (Fig. 3(b)). For
each time and field value, the signal is fitted as a function of \textgreek{b}
by $\delta\beta=F_{\theta}+G_{\phi}\cos2\beta+H_{\phi\varepsilon}\sin2\beta$.
The resulting fit is very good as seen in Fig. 3(b). The offset that
appears under the application of a magnetic field $\mu_{0}H=$6 mT
is related to the PMOKE signal $F_{\theta}=K\delta\theta$. The change
of the phase is a good indication of the presence of MLD. The resulting
$F_{\theta},\,G_{\phi},\,H_{\phi\varepsilon}$ time-dependent functions
are then fitted by sinusoidal fonctions at frequency $f_{SAW}$ in
order to extract the field-dependence of their amplitudes $f_{\theta},\,g_{\phi},\,h_{\phi\varepsilon}$,
respectively. We have $f_{\theta}=K\delta\theta_{0}(H)$ and $g_{\phi}=2V\delta\phi_{0}(H)\left|\cos2\phi_{0}(H)\right|$
where $\delta\theta_{0}$ and $\delta\phi_{0}$ are the amplitudes
of the oscillating $\delta\theta(t)$ and $\delta\phi(t)$, respectively.

\begin{figure}
\includegraphics[scale=0.35]{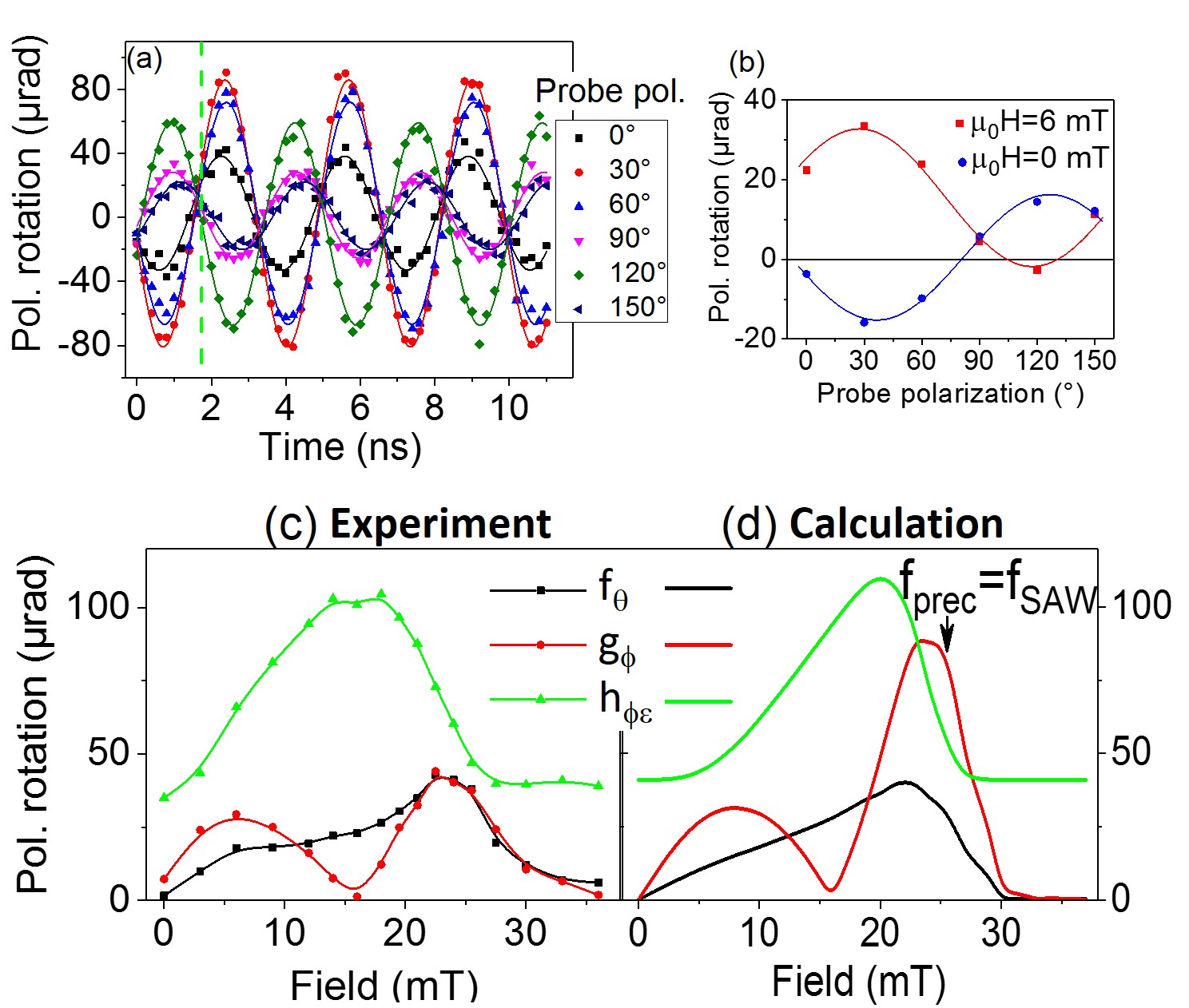}

\caption{Polarization and field dependence of the magnetization and strain
dynamics for $f_{SAW}=301$ MHz at $T$= 60 K. (a) TRPR signal for
six polarization orientations as a function of the optical delay for
$\mu_{0}H$= 6 mT. (b) TRPR at fixed time delay (cross section of
Fig. 3(a), green dashed line) versus probe polarization for two field
values. (c) Field variation of the three components $f_{\theta},\,g_{\phi},\,h_{\phi\varepsilon}$
related to polar Kerr effect (\textgreek{b}-independent), MLD (cos2\textgreek{b}
term) and mixed PE/MLD (sin2\textgreek{b} term), respectively, at
$f_{SAW}$=301 MHz. (d) Calculated amplitudes of $f_{\theta},\,g_{\phi},\,h_{\phi\varepsilon}$;
the parameters are $K$=18 mrad, $V$=0.8 mrad, $P_{E}$=0.77 rad
with $\varepsilon_{xx}=5.3\:10^{-5}$. The Gaussian distribution on
the in-plane anisotropy constant has a standard deviation equal to
12 \% of its mean value.}
\end{figure}

In Fig. 3(c) we plot the amplitudes $f_{\theta},\,g_{\phi},\,h_{\phi\varepsilon}$
as a function of the applied field. The amplitude of the PMOKE signal
$f_{\theta}$ (black curve) increases progressively to reach a maximum
at 22.5 mT and then drops to zero. The MLD component $g_{\phi}$ (red
curve) shows a different behavior with two maxima and a zero at $\mu_{0}H$=16
mT. The $h_{\phi\varepsilon}$ component reflecting a combination
of PE and MLD (green curve) has a broad maximum and a clear 40 \textmu rad
offset owing to the field-independent elasto-optic effect.

To demonstrate that the physics of the magneto-elastic coupling is
responsible for the experimental observations, the magnetization dynamics
is modelled in the framework of the Landau-Lifshitz-Gilbert equation
including the driving torque generated by the SAW \cite{Thevenard2013a}.
When the SAW is travelling along {[}1-10{]} the torque is mainly driven
by the $\varepsilon_{xx}$ strain component \cite{Kuszewski2018}.
We use the sample magnetic parameters (magnetization and magnetic
anisotropy) obtained from vibrating sample magnetometer and cavity-FMR
experiments. The calculated amplitudes \footnote{Because we are only considering the uniform FMR mode, it is unecessary
to take into account the absorption and optical phase shift of the
light in the layer, as in Ref. {[}\onlinecite{Shihab2017}{]} } of the magneto-optical signals $f_{\theta}$ and $g_{\phi}$ (Fig.
3(d), black and red curves, respectively) show a peak around 25 mT,
corresponding to the resonance condition of equal precession and SAW
frequencies at $f_{SAW}=301$ MHz. The $g_{\phi}$ amplitude (red
curve) goes to zero at a field such that the static magnetization
is at 45\textdegree{} of the easy axis $\left(\cos2\phi_{0}=0\right)$.
The baseline for $h_{\phi\varepsilon}$ is given by the PE effect
that does not depend on the magnetic field. As seen from the comparison
of Fig. 3(c) and (d) a good quantitative agreement between experimental
and calculated $f_{\theta},\,g_{\phi},\,h_{\phi\varepsilon}$ curves
is obtained. To account for the amplitude and width of $f_{\theta},\,g_{\phi},\,h_{\phi\varepsilon}$
we had to introduce a dispersion of the uniaxial in-plane magnetic
anisotropy constant that is known to be crucial for SAW-induced magnetization
dynamics \cite{Kuszewski2018}. These results show that we are able
to disentangle the different contributions in the optical polarization
signal and to accurately extract the dynamical magnetic contribution.
They provide the first direct, time-domain detection of magneto-acoustic
resonance induced by a \textit{propagating} SAW. Furthermore the presence
of Kerr and Voigt effects gives access to both the in-plane and out-of-plane
dynamical magnetization components.

\begin{figure*}
\includegraphics[width=17cm]{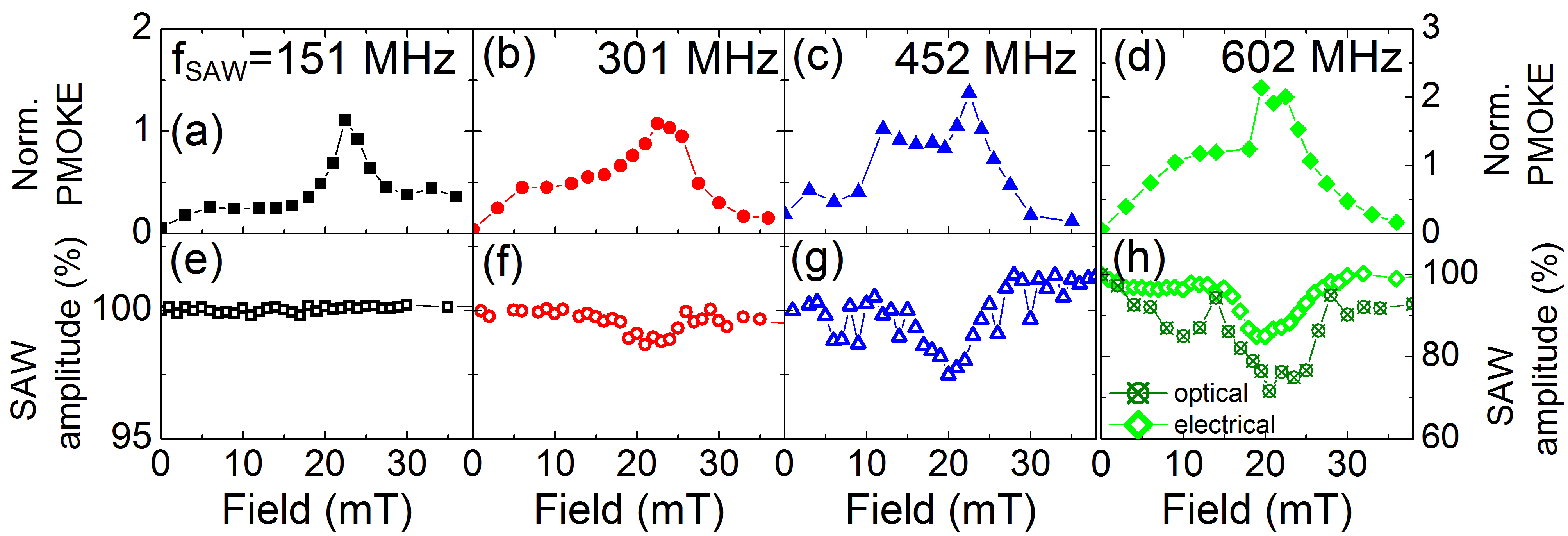}\caption{Top panels: Field dependence of the PMOKE signal (out-of-plane magnetization
dynamics) normalized by the field-independent photo-elastic signal
for four SAW frequencies at $T$=60 K. Bottom panels: variation of
the SAW amplitude with the magnetic field measured by acoustic-to-electrical
conversion at the receiving IDT and in (h) normalized PE signal detected
on GaAs in between the GaMnAs mesa and the receiving IDT (see crossed-circle
symbol in the schematics of Fig. 2), with a polarization angle $\beta$=45
deg. The amplitudes of excited SAW strain components are estimated
as $\varepsilon_{xx}=5.3\times10^{-5}$ and $\varepsilon_{zz}=1.8\times10^{-5}$
at $f_{SAW}$=301 MHz (they are slightly different at other frequencies
but low enough to stay in the linear regime).}
\end{figure*}

In order to compare the sensitivity of the magneto-optical signal
detection and the SAW amplitude detection after propagation along
the layer, we show in Fig. 4 the PMOKE component $f_{\theta}$ (out-of-plane
magnetization dynamics) and the variation of the electrically detected
SAW amplitude (through the receiving IDT) as a function of the applied
magnetic field for the four SAW frequencies. The PMOKE data (top panel
of Fig. 4) were normalized by the measured PE level (field-independent
baseline of $h_{\phi\varepsilon}$) in order to correct for slightly
different SAW amplitudes. For each frequency we observe a clear and
strong peak of the PMOKE curve at the resonance field (which has close
values for the four frequencies because of the steep variation of
the FMR frequency with the field \cite{Kuszewski2018}). The electrically
detected SAW amplitude (bottom panel of Fig. 4) is a rather noisy
signal at 151, 301 and 452 MHz, with a much smaller dynamical range.
At $f_{SAW}=$ 301, 452, and 602 MHz, it shows a dip at the same field
as the peak of the corresponding PMOKE signal. The dip is more pronounced
at the higher frequency of 602 MHz as expected from the increase of
the absorbed SAW power with the SAW frequency \cite{Dreher2012,Kuszewski2018}.
The dip is however not detectable at $f_{SAW}=151$ MHz whereas the
PMOKE peak is clearly detected. We also plot in Fig. 4(h) the amplitude
of the photo-elastic signal detected on GaAs, in between the GaMnAs
mesa and the receiving IDT (crossed-circle symbol in the scheme of
Fig. 2), normalized by its low-field value. The data was taken for
an incoming beam polarisation of 45 deg, which maximizes the strain-induced
birefringence. Its field-dependence is similar to that of the electrical
signal (Fig. 4(h)) in shape and amplitude. These results show the
much larger dynamical range and sensitivity of the magneto-optical
SAW-FMR signal, because it is detected on a zero background, with
respect to electrical or optical signal of the SAW amplitude variation,
detected on a non-zero background.

\section{Conclusion}

We have developed a sensitive time-domain optical technique to investigate
the magneto-elastic coupling between piezo-electrically generated
travelling SAWs and magnetization in ferromagnetic layers at variable
temperature. The time-resolved magnetization precession was clearly
proved to originate from the magneto-elastic coupling with the SAW
and shows a resonant behavior at equal SAW and precession frequencies.
Compared to the detection of the SAW attenuation, the magneto-optical
time-resolved SAW-FMR signal is more sensitive and provides a better
signal-to-noise ratio. The detection threshold of the time-resolved
signal is mainly governed by the convolution of the SAW wavelength
and the laser spot size $w$. As such, magnetization precession could
still be observed at $f_{SAW}$=900 MHz, for which for $\lambda_{SAW}\sim3w$.
Working at higher SAW frequencies (above 1 GHz) will require a tighter
focusing of the laser spot, either by working at higher photon energy
(keeping a decent Kerr signal) and/or by increasing the numerical
aperture of the focusing objective (keeping a good linear polarization).
More generally, we believe that this approach, combining high space/time
sensitivity and access to the two components of magnetization will
find a wider use in any experiment requiring the synchronization of
a radio-frequency electrical stimulus with the ultra-fast optical
detection of the magnetic effects it induces, and should benefit broadly
the magnetization dynamics community. Applied to magneto-strictive
materials, this technique opens the way for a deeper insight into
the magnon-phonon coupling and the exploration of the nonlinear regime
of acoustic-wave induced magnetization dynamics. 
\begin{acknowledgments}
This work has been supported by the French Agence Nationale de la
Recherche (ANR13-JS04-0001-01). The authors also acknowledge D. Hrabovsky
(MPBT-Physical Properties Low Temperature facility of Sorbonne Université),
H. J. von Bardeleben (INSP) for magnetometry measurements, and C.
Testelin (INSP) for providing scientific equipment. 
\end{acknowledgments}

 \bibliographystyle{apsrev4-1}
\bibliography{library}

\begin{thebibliography}{37}%
\makeatletter
\providecommand \@ifxundefined [1]{%
 \@ifx{#1\undefined}
}%
\providecommand \@ifnum [1]{%
 \ifnum #1\expandafter \@firstoftwo
 \else \expandafter \@secondoftwo
 \fi
}%
\providecommand \@ifx [1]{%
 \ifx #1\expandafter \@firstoftwo
 \else \expandafter \@secondoftwo
 \fi
}%
\providecommand \natexlab [1]{#1}%
\providecommand \enquote  [1]{``#1''}%
\providecommand \bibnamefont  [1]{#1}%
\providecommand \bibfnamefont [1]{#1}%
\providecommand \citenamefont [1]{#1}%
\providecommand \href@noop [0]{\@secondoftwo}%
\providecommand \href [0]{\begingroup \@sanitize@url \@href}%
\providecommand \@href[1]{\@@startlink{#1}\@@href}%
\providecommand \@@href[1]{\endgroup#1\@@endlink}%
\providecommand \@sanitize@url [0]{\catcode `\\12\catcode `\$12\catcode
  `\&12\catcode `\#12\catcode `\^12\catcode `\_12\catcode `\%12\relax}%
\providecommand \@@startlink[1]{}%
\providecommand \@@endlink[0]{}%
\providecommand \url  [0]{\begingroup\@sanitize@url \@url }%
\providecommand \@url [1]{\endgroup\@href {#1}{\urlprefix }}%
\providecommand \urlprefix  [0]{URL }%
\providecommand \Eprint [0]{\href }%
\providecommand \doibase [0]{http://dx.doi.org/}%
\providecommand \selectlanguage [0]{\@gobble}%
\providecommand \bibinfo  [0]{\@secondoftwo}%
\providecommand \bibfield  [0]{\@secondoftwo}%
\providecommand \translation [1]{[#1]}%
\providecommand \BibitemOpen [0]{}%
\providecommand \bibitemStop [0]{}%
\providecommand \bibitemNoStop [0]{.\EOS\space}%
\providecommand \EOS [0]{\spacefactor3000\relax}%
\providecommand \BibitemShut  [1]{\csname bibitem#1\endcsname}%
\let\auto@bib@innerbib\@empty
\bibitem [{\citenamefont {Dreher}\ \emph {et~al.}(2012)\citenamefont {Dreher},
  \citenamefont {Weiler}, \citenamefont {Pernpeintner}, \citenamefont {Huebl},
  \citenamefont {Gross}, \citenamefont {Brandt},\ and\ \citenamefont
  {Goennenwein}}]{Dreher2012}%
  \BibitemOpen
  \bibfield  {author} {\bibinfo {author} {\bibfnamefont {L.}~\bibnamefont
  {Dreher}}, \bibinfo {author} {\bibfnamefont {M.}~\bibnamefont {Weiler}},
  \bibinfo {author} {\bibfnamefont {M.}~\bibnamefont {Pernpeintner}}, \bibinfo
  {author} {\bibfnamefont {H.}~\bibnamefont {Huebl}}, \bibinfo {author}
  {\bibfnamefont {R.}~\bibnamefont {Gross}}, \bibinfo {author} {\bibfnamefont
  {M.~S.}\ \bibnamefont {Brandt}}, \ and\ \bibinfo {author} {\bibfnamefont
  {S.~T.~B.}\ \bibnamefont {Goennenwein}},\ }\href {\doibase
  10.1103/PhysRevB.86.134415} {\bibfield  {journal} {\bibinfo  {journal} {Phys.
  Rev. B}\ }\textbf {\bibinfo {volume} {86}},\ \bibinfo {pages} {134415}
  (\bibinfo {year} {2012})}\BibitemShut {NoStop}%
\bibitem [{\citenamefont {Bombeck}\ \emph {et~al.}(2012)\citenamefont
  {Bombeck}, \citenamefont {Salasyuk}, \citenamefont {Glavin}, \citenamefont
  {Scherbakov}, \citenamefont {Br{\"{u}}ggemann}, \citenamefont {Yakovlev},
  \citenamefont {Sapega}, \citenamefont {Liu}, \citenamefont {Furdyna},
  \citenamefont {Akimov},\ and\ \citenamefont {Bayer}}]{Bombeck2012}%
  \BibitemOpen
  \bibfield  {author} {\bibinfo {author} {\bibfnamefont {M.}~\bibnamefont
  {Bombeck}}, \bibinfo {author} {\bibfnamefont {A.~S.}\ \bibnamefont
  {Salasyuk}}, \bibinfo {author} {\bibfnamefont {B.~A.}\ \bibnamefont
  {Glavin}}, \bibinfo {author} {\bibfnamefont {A.~V.}\ \bibnamefont
  {Scherbakov}}, \bibinfo {author} {\bibfnamefont {C.}~\bibnamefont
  {Br{\"{u}}ggemann}}, \bibinfo {author} {\bibfnamefont {D.~R.}\ \bibnamefont
  {Yakovlev}}, \bibinfo {author} {\bibfnamefont {V.~F.}\ \bibnamefont
  {Sapega}}, \bibinfo {author} {\bibfnamefont {X.}~\bibnamefont {Liu}},
  \bibinfo {author} {\bibfnamefont {J.~K.}\ \bibnamefont {Furdyna}}, \bibinfo
  {author} {\bibfnamefont {A.~V.}\ \bibnamefont {Akimov}}, \ and\ \bibinfo
  {author} {\bibfnamefont {M.}~\bibnamefont {Bayer}},\ }\href {\doibase
  10.1103/PhysRevB.85.195324} {\bibfield  {journal} {\bibinfo  {journal} {Phys.
  Rev. B}\ }\textbf {\bibinfo {volume} {85}},\ \bibinfo {pages} {195324}
  (\bibinfo {year} {2012})}\BibitemShut {NoStop}%
\bibitem [{\citenamefont {Li}\ \emph {et~al.}(2014)\citenamefont {Li},
  \citenamefont {Buford}, \citenamefont {Jander},\ and\ \citenamefont
  {Dhagat}}]{Li2014b}%
  \BibitemOpen
  \bibfield  {author} {\bibinfo {author} {\bibfnamefont {W.}~\bibnamefont
  {Li}}, \bibinfo {author} {\bibfnamefont {B.}~\bibnamefont {Buford}}, \bibinfo
  {author} {\bibfnamefont {A.}~\bibnamefont {Jander}}, \ and\ \bibinfo {author}
  {\bibfnamefont {P.}~\bibnamefont {Dhagat}},\ }\href {\doibase
  10.1109/TMAG.2013.2285018} {\bibfield  {journal} {\bibinfo  {journal} {IEEE
  Transactions on Magnetics}\ }\textbf {\bibinfo {volume} {50}},\ \bibinfo
  {pages} {37} (\bibinfo {year} {2014})}\BibitemShut {NoStop}%
\bibitem [{\citenamefont {Thevenard}\ \emph {et~al.}(2014)\citenamefont
  {Thevenard}, \citenamefont {Gourdon}, \citenamefont {Prieur}, \citenamefont
  {von Bardeleben}, \citenamefont {Vincent}, \citenamefont {Becerra},
  \citenamefont {Largeau},\ and\ \citenamefont {Duquesne}}]{Thevenard2014}%
  \BibitemOpen
  \bibfield  {author} {\bibinfo {author} {\bibfnamefont {L.}~\bibnamefont
  {Thevenard}}, \bibinfo {author} {\bibfnamefont {C.}~\bibnamefont {Gourdon}},
  \bibinfo {author} {\bibfnamefont {J.~Y.}\ \bibnamefont {Prieur}}, \bibinfo
  {author} {\bibfnamefont {H.~J.}\ \bibnamefont {von Bardeleben}}, \bibinfo
  {author} {\bibfnamefont {S.}~\bibnamefont {Vincent}}, \bibinfo {author}
  {\bibfnamefont {L.}~\bibnamefont {Becerra}}, \bibinfo {author} {\bibfnamefont
  {L.}~\bibnamefont {Largeau}}, \ and\ \bibinfo {author} {\bibfnamefont
  {J.-Y.}\ \bibnamefont {Duquesne}},\ }\href {\doibase
  10.1103/PhysRevB.90.094401} {\bibfield  {journal} {\bibinfo  {journal} {Phys.
  Rev. B}\ }\textbf {\bibinfo {volume} {90}},\ \bibinfo {pages} {094401}
  (\bibinfo {year} {2014})}\BibitemShut {NoStop}%
\bibitem [{\citenamefont {Yahagi}\ \emph {et~al.}(2014)\citenamefont {Yahagi},
  \citenamefont {Harteneck}, \citenamefont {Cabrini},\ and\ \citenamefont
  {Schmidt}}]{Yahagi2014}%
  \BibitemOpen
  \bibfield  {author} {\bibinfo {author} {\bibfnamefont {Y.}~\bibnamefont
  {Yahagi}}, \bibinfo {author} {\bibfnamefont {B.}~\bibnamefont {Harteneck}},
  \bibinfo {author} {\bibfnamefont {S.}~\bibnamefont {Cabrini}}, \ and\
  \bibinfo {author} {\bibfnamefont {H.}~\bibnamefont {Schmidt}},\ }\href
  {\doibase 10.1103/PhysRevB.90.140405} {\bibfield  {journal} {\bibinfo
  {journal} {Physical Review B}\ }\textbf {\bibinfo {volume} {90}},\ \bibinfo
  {pages} {140405} (\bibinfo {year} {2014})}\BibitemShut {NoStop}%
\bibitem [{\citenamefont {Shen}\ and\ \citenamefont {Bauer}(2015)}]{Shen2015a}%
  \BibitemOpen
  \bibfield  {author} {\bibinfo {author} {\bibfnamefont {K.}~\bibnamefont
  {Shen}}\ and\ \bibinfo {author} {\bibfnamefont {G.~E.}\ \bibnamefont
  {Bauer}},\ }\href {\doibase 10.1103/PhysRevLett.115.197201} {\bibfield
  {journal} {\bibinfo  {journal} {Phys. Rev. Lett.}\ }\textbf {\bibinfo
  {volume} {115}},\ \bibinfo {pages} {197201} (\bibinfo {year}
  {2015})}\BibitemShut {NoStop}%
\bibitem [{\citenamefont {Davis}\ \emph {et~al.}(2015)\citenamefont {Davis},
  \citenamefont {Borchers}, \citenamefont {Maranville},\ and\ \citenamefont
  {Adenwalla}}]{Davis2015}%
  \BibitemOpen
  \bibfield  {author} {\bibinfo {author} {\bibfnamefont {S.}~\bibnamefont
  {Davis}}, \bibinfo {author} {\bibfnamefont {J.~a.}\ \bibnamefont {Borchers}},
  \bibinfo {author} {\bibfnamefont {B.~B.}\ \bibnamefont {Maranville}}, \ and\
  \bibinfo {author} {\bibfnamefont {S.}~\bibnamefont {Adenwalla}},\ }\href
  {\doibase 10.1063/1.4907580} {\bibfield  {journal} {\bibinfo  {journal}
  {Journal of Applied Physics}\ }\textbf {\bibinfo {volume} {117}},\ \bibinfo
  {pages} {063904} (\bibinfo {year} {2015})}\BibitemShut {NoStop}%
\bibitem [{\citenamefont {Gowtham}\ \emph {et~al.}(2015)\citenamefont
  {Gowtham}, \citenamefont {Moriyama}, \citenamefont {Ralph},\ and\
  \citenamefont {Buhrman}}]{Gowtham2015}%
  \BibitemOpen
  \bibfield  {author} {\bibinfo {author} {\bibfnamefont {P.~G.}\ \bibnamefont
  {Gowtham}}, \bibinfo {author} {\bibfnamefont {T.}~\bibnamefont {Moriyama}},
  \bibinfo {author} {\bibfnamefont {D.~C.}\ \bibnamefont {Ralph}}, \ and\
  \bibinfo {author} {\bibfnamefont {R.~A.}\ \bibnamefont {Buhrman}},\ }\href
  {\doibase 10.1063/1.4938390} {\bibfield  {journal} {\bibinfo  {journal}
  {Journal of Applied Physics}\ }\textbf {\bibinfo {volume} {118}},\ \bibinfo
  {pages} {233910} (\bibinfo {year} {2015})}\BibitemShut {NoStop}%
\bibitem [{\citenamefont {Chowdhury}\ \emph {et~al.}(2015)\citenamefont
  {Chowdhury}, \citenamefont {Dhagat},\ and\ \citenamefont
  {Jander}}]{Chowdhury2015}%
  \BibitemOpen
  \bibfield  {author} {\bibinfo {author} {\bibfnamefont {P.}~\bibnamefont
  {Chowdhury}}, \bibinfo {author} {\bibfnamefont {P.}~\bibnamefont {Dhagat}}, \
  and\ \bibinfo {author} {\bibfnamefont {A.}~\bibnamefont {Jander}},\ }\href
  {\doibase 10.1109/TMAG.2015.2445791} {\bibfield  {journal} {\bibinfo
  {journal} {IEEE Transactions on Magnetics}\ }\textbf {\bibinfo {volume}
  {51}},\ \bibinfo {pages} {1300904} (\bibinfo {year} {2015})}\BibitemShut
  {NoStop}%
\bibitem [{\citenamefont {Janu{\v{s}}onis}\ \emph {et~al.}(2015)\citenamefont
  {Janu{\v{s}}onis}, \citenamefont {Chang}, \citenamefont {van Loosdrecht},\
  and\ \citenamefont {Tobey}}]{Janusonis2015a}%
  \BibitemOpen
  \bibfield  {author} {\bibinfo {author} {\bibfnamefont {J.}~\bibnamefont
  {Janu{\v{s}}onis}}, \bibinfo {author} {\bibfnamefont {C.~L.}\ \bibnamefont
  {Chang}}, \bibinfo {author} {\bibfnamefont {P.~H.~M.}\ \bibnamefont {van
  Loosdrecht}}, \ and\ \bibinfo {author} {\bibfnamefont {R.~I.}\ \bibnamefont
  {Tobey}},\ }\href {\doibase 10.1063/1.4919882} {\bibfield  {journal}
  {\bibinfo  {journal} {Applied Physics Letters}\ }\textbf {\bibinfo {volume}
  {106}},\ \bibinfo {pages} {181601} (\bibinfo {year} {2015})}\BibitemShut
  {NoStop}%
\bibitem [{\citenamefont {Thevenard}\ \emph
  {et~al.}(2016{\natexlab{a}})\citenamefont {Thevenard}, \citenamefont
  {Camara}, \citenamefont {Prieur}, \citenamefont {Rovillain}, \citenamefont
  {Lema{\^{i}}tre}, \citenamefont {Gourdon},\ and\ \citenamefont
  {Duquesne}}]{Thevenard2016a}%
  \BibitemOpen
  \bibfield  {author} {\bibinfo {author} {\bibfnamefont {L.}~\bibnamefont
  {Thevenard}}, \bibinfo {author} {\bibfnamefont {I.~S.}\ \bibnamefont
  {Camara}}, \bibinfo {author} {\bibfnamefont {J.-Y.}\ \bibnamefont {Prieur}},
  \bibinfo {author} {\bibfnamefont {P.}~\bibnamefont {Rovillain}}, \bibinfo
  {author} {\bibfnamefont {A.}~\bibnamefont {Lema{\^{i}}tre}}, \bibinfo
  {author} {\bibfnamefont {C.}~\bibnamefont {Gourdon}}, \ and\ \bibinfo
  {author} {\bibfnamefont {J.-Y.}\ \bibnamefont {Duquesne}},\ }\href {\doibase
  10.1103/PhysRevB.93.140405} {\bibfield  {journal} {\bibinfo  {journal}
  {Physical Review B}\ }\textbf {\bibinfo {volume} {93}},\ \bibinfo {pages}
  {140405} (\bibinfo {year} {2016}{\natexlab{a}})}\BibitemShut {NoStop}%
\bibitem [{\citenamefont {Thevenard}\ \emph
  {et~al.}(2016{\natexlab{b}})\citenamefont {Thevenard}, \citenamefont
  {Camara}, \citenamefont {Majrab}, \citenamefont {Bernard}, \citenamefont
  {Rovillain}, \citenamefont {Lema{\^{i}}tre}, \citenamefont {Gourdon},\ and\
  \citenamefont {Duquesne}}]{Thevenard2016}%
  \BibitemOpen
  \bibfield  {author} {\bibinfo {author} {\bibfnamefont {L.}~\bibnamefont
  {Thevenard}}, \bibinfo {author} {\bibfnamefont {I.~S.}\ \bibnamefont
  {Camara}}, \bibinfo {author} {\bibfnamefont {S.}~\bibnamefont {Majrab}},
  \bibinfo {author} {\bibfnamefont {M.}~\bibnamefont {Bernard}}, \bibinfo
  {author} {\bibfnamefont {P.}~\bibnamefont {Rovillain}}, \bibinfo {author}
  {\bibfnamefont {A.}~\bibnamefont {Lema{\^{i}}tre}}, \bibinfo {author}
  {\bibfnamefont {C.}~\bibnamefont {Gourdon}}, \ and\ \bibinfo {author}
  {\bibfnamefont {J.-Y.}\ \bibnamefont {Duquesne}},\ }\href {\doibase
  10.1103/PhysRevB.93.134430} {\bibfield  {journal} {\bibinfo  {journal}
  {Physical Review B}\ }\textbf {\bibinfo {volume} {93}},\ \bibinfo {pages}
  {134430} (\bibinfo {year} {2016}{\natexlab{b}})}\BibitemShut {NoStop}%
\bibitem [{\citenamefont {Chang}\ \emph {et~al.}(2017)\citenamefont {Chang},
  \citenamefont {Lomonosov}, \citenamefont {Janusonis}, \citenamefont {Vlasov},
  \citenamefont {Temnov},\ and\ \citenamefont {Tobey}}]{Chang2017a}%
  \BibitemOpen
  \bibfield  {author} {\bibinfo {author} {\bibfnamefont {C.~L.}\ \bibnamefont
  {Chang}}, \bibinfo {author} {\bibfnamefont {A.~M.}\ \bibnamefont
  {Lomonosov}}, \bibinfo {author} {\bibfnamefont {J.}~\bibnamefont
  {Janusonis}}, \bibinfo {author} {\bibfnamefont {V.~S.}\ \bibnamefont
  {Vlasov}}, \bibinfo {author} {\bibfnamefont {V.~V.}\ \bibnamefont {Temnov}},
  \ and\ \bibinfo {author} {\bibfnamefont {R.~I.}\ \bibnamefont {Tobey}},\
  }\href {\doibase 10.1103/PhysRevB.95.060409} {\bibfield  {journal} {\bibinfo
  {journal} {Phys. Rev. B}\ }\textbf {\bibinfo {volume} {95}},\ \bibinfo
  {pages} {060409} (\bibinfo {year} {2017})}\BibitemShut {NoStop}%
\bibitem [{\citenamefont {Hashimoto}\ \emph {et~al.}(2017)\citenamefont
  {Hashimoto}, \citenamefont {Daimon}, \citenamefont {Iguchi}, \citenamefont
  {Oikawa}, \citenamefont {Shen}, \citenamefont {Sato}, \citenamefont
  {Bossini}, \citenamefont {Tabuchi}, \citenamefont {Satoh}, \citenamefont
  {Hillebrands}, \citenamefont {Bauer}, \citenamefont {Johansen}, \citenamefont
  {Kirilyuk}, \citenamefont {Rasing},\ and\ \citenamefont
  {Saitoh}}]{Hashimoto2017a}%
  \BibitemOpen
  \bibfield  {author} {\bibinfo {author} {\bibfnamefont {Y.}~\bibnamefont
  {Hashimoto}}, \bibinfo {author} {\bibfnamefont {S.}~\bibnamefont {Daimon}},
  \bibinfo {author} {\bibfnamefont {R.}~\bibnamefont {Iguchi}}, \bibinfo
  {author} {\bibfnamefont {Y.}~\bibnamefont {Oikawa}}, \bibinfo {author}
  {\bibfnamefont {K.}~\bibnamefont {Shen}}, \bibinfo {author} {\bibfnamefont
  {K.}~\bibnamefont {Sato}}, \bibinfo {author} {\bibfnamefont {D.}~\bibnamefont
  {Bossini}}, \bibinfo {author} {\bibfnamefont {Y.}~\bibnamefont {Tabuchi}},
  \bibinfo {author} {\bibfnamefont {T.}~\bibnamefont {Satoh}}, \bibinfo
  {author} {\bibfnamefont {B.}~\bibnamefont {Hillebrands}}, \bibinfo {author}
  {\bibfnamefont {G.~E.}\ \bibnamefont {Bauer}}, \bibinfo {author}
  {\bibfnamefont {T.~H.}\ \bibnamefont {Johansen}}, \bibinfo {author}
  {\bibfnamefont {A.}~\bibnamefont {Kirilyuk}}, \bibinfo {author}
  {\bibfnamefont {T.}~\bibnamefont {Rasing}}, \ and\ \bibinfo {author}
  {\bibfnamefont {E.}~\bibnamefont {Saitoh}},\ }\href {\doibase
  10.1038/ncomms15859} {\bibfield  {journal} {\bibinfo  {journal} {Nat.
  Commun.}\ }\textbf {\bibinfo {volume} {8}},\ \bibinfo {pages} {15859}
  (\bibinfo {year} {2017})}\BibitemShut {NoStop}%
\bibitem [{\citenamefont {Kim}\ and\ \citenamefont {Bigot}(2017)}]{Kim2017}%
  \BibitemOpen
  \bibfield  {author} {\bibinfo {author} {\bibfnamefont {J.-W.}\ \bibnamefont
  {Kim}}\ and\ \bibinfo {author} {\bibfnamefont {J.-Y.}\ \bibnamefont
  {Bigot}},\ }\href {\doibase 10.1103/PhysRevB.95.144422} {\bibfield  {journal}
  {\bibinfo  {journal} {Physical Review B}\ }\textbf {\bibinfo {volume} {95}},\
  \bibinfo {pages} {144422} (\bibinfo {year} {2017})}\BibitemShut {NoStop}%
\bibitem [{\citenamefont {Kuszewski}\ \emph {et~al.}(2018)\citenamefont
  {Kuszewski}, \citenamefont {Camara}, \citenamefont {Biarrotte}, \citenamefont
  {Becerra}, \citenamefont {von Bardeleben}, \citenamefont {{Savero Torres}},
  \citenamefont {Lema{\^{i}}tre}, \citenamefont {Gourdon}, \citenamefont
  {Duquesne},\ and\ \citenamefont {Thevenard}}]{Kuszewski2018}%
  \BibitemOpen
  \bibfield  {author} {\bibinfo {author} {\bibfnamefont {P.}~\bibnamefont
  {Kuszewski}}, \bibinfo {author} {\bibfnamefont {I.~S.}\ \bibnamefont
  {Camara}}, \bibinfo {author} {\bibfnamefont {N.}~\bibnamefont {Biarrotte}},
  \bibinfo {author} {\bibfnamefont {L.}~\bibnamefont {Becerra}}, \bibinfo
  {author} {\bibfnamefont {J.}~\bibnamefont {von Bardeleben}}, \bibinfo
  {author} {\bibfnamefont {W.}~\bibnamefont {{Savero Torres}}}, \bibinfo
  {author} {\bibfnamefont {A.}~\bibnamefont {Lema{\^{i}}tre}}, \bibinfo
  {author} {\bibfnamefont {C.}~\bibnamefont {Gourdon}}, \bibinfo {author}
  {\bibfnamefont {J.-Y.}\ \bibnamefont {Duquesne}}, \ and\ \bibinfo {author}
  {\bibfnamefont {L.}~\bibnamefont {Thevenard}},\ }\href {\doibase
  10.1088/1361-648X/aac152} {\bibfield  {journal} {\bibinfo  {journal} {Journal
  of Physics: Condensed Matter}\ }\textbf {\bibinfo {volume} {30}},\ \bibinfo
  {pages} {244003} (\bibinfo {year} {2018})}\BibitemShut {NoStop}%
\bibitem [{\citenamefont {Lema\^{i}tre}\ \emph {et~al.}(2008)\citenamefont
  {Lema\^{i}tre}, \citenamefont {Miard}, \citenamefont {Travers}, \citenamefont
  {Mauguin}, \citenamefont {Largeau}, \citenamefont {Gourdon}, \citenamefont
  {Jeudy}, \citenamefont {Tran},\ and\ \citenamefont {George}}]{Lemaitre2008}%
  \BibitemOpen
  \bibfield  {author} {\bibinfo {author} {\bibfnamefont {A.}~\bibnamefont
  {Lema\^{i}tre}}, \bibinfo {author} {\bibfnamefont {A.}~\bibnamefont {Miard}},
  \bibinfo {author} {\bibfnamefont {L.}~\bibnamefont {Travers}}, \bibinfo
  {author} {\bibfnamefont {O.}~\bibnamefont {Mauguin}}, \bibinfo {author}
  {\bibfnamefont {L.}~\bibnamefont {Largeau}}, \bibinfo {author} {\bibfnamefont
  {C.}~\bibnamefont {Gourdon}}, \bibinfo {author} {\bibfnamefont
  {V.}~\bibnamefont {Jeudy}}, \bibinfo {author} {\bibfnamefont
  {M.}~\bibnamefont {Tran}}, \ and\ \bibinfo {author} {\bibfnamefont {J.-M.}\
  \bibnamefont {George}},\ }\href {\doibase 10.1063/1.2963979} {\bibfield
  {journal} {\bibinfo  {journal} {Applied Physics Letters}\ }\textbf {\bibinfo
  {volume} {93}},\ \bibinfo {pages} {021123} (\bibinfo {year}
  {2008})}\BibitemShut {NoStop}%
\bibitem [{\citenamefont {Nath}\ \emph {et~al.}(1999)\citenamefont {Nath},
  \citenamefont {Rao}, \citenamefont {Lavric}, \citenamefont {Eom},
  \citenamefont {Wu},\ and\ \citenamefont {Tsui}}]{Nath1999}%
  \BibitemOpen
  \bibfield  {author} {\bibinfo {author} {\bibfnamefont {T.~K.}\ \bibnamefont
  {Nath}}, \bibinfo {author} {\bibfnamefont {R.~A.}\ \bibnamefont {Rao}},
  \bibinfo {author} {\bibfnamefont {D.}~\bibnamefont {Lavric}}, \bibinfo
  {author} {\bibfnamefont {C.~B.}\ \bibnamefont {Eom}}, \bibinfo {author}
  {\bibfnamefont {L.}~\bibnamefont {Wu}}, \ and\ \bibinfo {author}
  {\bibfnamefont {F.}~\bibnamefont {Tsui}},\ }\href {\doibase 10.1063/1.123634}
  {\bibfield  {journal} {\bibinfo  {journal} {Appl. Phys. Lett.}\ }\textbf
  {\bibinfo {volume} {74}},\ \bibinfo {pages} {1615} (\bibinfo {year}
  {1999})}\BibitemShut {NoStop}%
\bibitem [{\citenamefont {Linnik}\ \emph {et~al.}(2011)\citenamefont {Linnik},
  \citenamefont {Scherbakov}, \citenamefont {Yakovlev}, \citenamefont {Liu},
  \citenamefont {Furdyna},\ and\ \citenamefont {Bayer}}]{Linnik2011}%
  \BibitemOpen
  \bibfield  {author} {\bibinfo {author} {\bibfnamefont {T.~L.}\ \bibnamefont
  {Linnik}}, \bibinfo {author} {\bibfnamefont {A.~V.}\ \bibnamefont
  {Scherbakov}}, \bibinfo {author} {\bibfnamefont {D.~R.}\ \bibnamefont
  {Yakovlev}}, \bibinfo {author} {\bibfnamefont {X.}~\bibnamefont {Liu}},
  \bibinfo {author} {\bibfnamefont {J.~K.}\ \bibnamefont {Furdyna}}, \ and\
  \bibinfo {author} {\bibfnamefont {M.}~\bibnamefont {Bayer}},\ }\href
  {\doibase 10.1103/PhysRevB.84.214432} {\bibfield  {journal} {\bibinfo
  {journal} {Physical Review B}\ }\textbf {\bibinfo {volume} {84}},\ \bibinfo
  {pages} {214432} (\bibinfo {year} {2011})}\BibitemShut {NoStop}%
\bibitem [{\citenamefont {Thevenard}\ \emph {et~al.}(2013)\citenamefont
  {Thevenard}, \citenamefont {Duquesne}, \citenamefont {Peronne}, \citenamefont
  {von Bardeleben}, \citenamefont {Jaffres}, \citenamefont {Ruttala},
  \citenamefont {George}, \citenamefont {Lema{\^{\i}}tre},\ and\ \citenamefont
  {Gourdon}}]{Thevenard2013a}%
  \BibitemOpen
  \bibfield  {author} {\bibinfo {author} {\bibfnamefont {L.}~\bibnamefont
  {Thevenard}}, \bibinfo {author} {\bibfnamefont {J.-Y.}\ \bibnamefont
  {Duquesne}}, \bibinfo {author} {\bibfnamefont {E.}~\bibnamefont {Peronne}},
  \bibinfo {author} {\bibfnamefont {H.~J.}\ \bibnamefont {von Bardeleben}},
  \bibinfo {author} {\bibfnamefont {H.}~\bibnamefont {Jaffres}}, \bibinfo
  {author} {\bibfnamefont {S.}~\bibnamefont {Ruttala}}, \bibinfo {author}
  {\bibfnamefont {J.-M.}\ \bibnamefont {George}}, \bibinfo {author}
  {\bibfnamefont {A.}~\bibnamefont {Lema{\^{\i}}tre}}, \ and\ \bibinfo {author}
  {\bibfnamefont {C.}~\bibnamefont {Gourdon}},\ }\href {\doibase
  10.1103/PhysRevB.87.144402} {\bibfield  {journal} {\bibinfo  {journal} {Phys.
  Rev. B}\ }\textbf {\bibinfo {volume} {87}},\ \bibinfo {pages} {144402}
  (\bibinfo {year} {2013})}\BibitemShut {NoStop}%
\bibitem [{\citenamefont {Yeo}\ and\ \citenamefont {Friend}(2014)}]{Yeo2014}%
  \BibitemOpen
  \bibfield  {author} {\bibinfo {author} {\bibfnamefont {L.~Y.}\ \bibnamefont
  {Yeo}}\ and\ \bibinfo {author} {\bibfnamefont {J.~R.}\ \bibnamefont
  {Friend}},\ }\href {\doibase 10.1146/annurev-fluid-010313-141418} {\bibfield
  {journal} {\bibinfo  {journal} {Annu. Rev. Fluid Mech.}\ }\textbf {\bibinfo
  {volume} {46}},\ \bibinfo {pages} {379} (\bibinfo {year} {2014})}\BibitemShut
  {NoStop}%
\bibitem [{\citenamefont {Mamishev}\ \emph {et~al.}(2004)\citenamefont
  {Mamishev}, \citenamefont {Sundara-Rajan}, \citenamefont {Yang},
  \citenamefont {Du},\ and\ \citenamefont {Zahn}}]{Mamishev2004}%
  \BibitemOpen
  \bibfield  {author} {\bibinfo {author} {\bibfnamefont {A.~V.}\ \bibnamefont
  {Mamishev}}, \bibinfo {author} {\bibfnamefont {K.}~\bibnamefont
  {Sundara-Rajan}}, \bibinfo {author} {\bibfnamefont {F.}~\bibnamefont {Yang}},
  \bibinfo {author} {\bibfnamefont {Y.}~\bibnamefont {Du}}, \ and\ \bibinfo
  {author} {\bibfnamefont {M.}~\bibnamefont {Zahn}},\ }\href {\doibase
  10.1109/JPROC.2004.826603} {\bibfield  {journal} {\bibinfo  {journal} {Proc.
  IEEE}\ }\textbf {\bibinfo {volume} {92}},\ \bibinfo {pages} {808} (\bibinfo
  {year} {2004})}\BibitemShut {NoStop}%
\bibitem [{\citenamefont {Poole}\ and\ \citenamefont {Nash}(2017)}]{Poole2017}%
  \BibitemOpen
  \bibfield  {author} {\bibinfo {author} {\bibfnamefont {T.}~\bibnamefont
  {Poole}}\ and\ \bibinfo {author} {\bibfnamefont {G.~R.}\ \bibnamefont
  {Nash}},\ }\href {\doibase 10.1038/s41598-017-01979-8} {\bibfield  {journal}
  {\bibinfo  {journal} {Sci. Rep.}\ }\textbf {\bibinfo {volume} {7}},\ \bibinfo
  {pages} {1767} (\bibinfo {year} {2017})}\BibitemShut {NoStop}%
\bibitem [{\citenamefont {{De Lima}}\ and\ \citenamefont
  {Santos}(2005)}]{DeLima2005}%
  \BibitemOpen
  \bibfield  {author} {\bibinfo {author} {\bibfnamefont {M.~M.}\ \bibnamefont
  {{De Lima}}}\ and\ \bibinfo {author} {\bibfnamefont {P.~V.}\ \bibnamefont
  {Santos}},\ }\href {\doibase 10.1088/0034-4885/68/7/R02} {\bibfield
  {journal} {\bibinfo  {journal} {Reports Prog. Phys.}\ }\textbf {\bibinfo
  {volume} {68}},\ \bibinfo {pages} {1639} (\bibinfo {year}
  {2005})}\BibitemShut {NoStop}%
\bibitem [{\citenamefont {Neubrand}\ and\ \citenamefont
  {Hess}(1992)}]{Neubrand1992}%
  \BibitemOpen
  \bibfield  {author} {\bibinfo {author} {\bibfnamefont {A.}~\bibnamefont
  {Neubrand}}\ and\ \bibinfo {author} {\bibfnamefont {P.}~\bibnamefont
  {Hess}},\ }\href {\doibase 10.1063/1.350747} {\bibfield  {journal} {\bibinfo
  {journal} {Journal of Applied Physics}\ }\textbf {\bibinfo {volume} {71}},\
  \bibinfo {pages} {227} (\bibinfo {year} {1992})}\BibitemShut {NoStop}%
\bibitem [{\citenamefont {Royer}\ and\ \citenamefont
  {Dieulesaint}(2000{\natexlab{a}})}]{Royer2}%
  \BibitemOpen
  \bibfield  {author} {\bibinfo {author} {\bibfnamefont {D.}~\bibnamefont
  {Royer}}\ and\ \bibinfo {author} {\bibfnamefont {E.}~\bibnamefont
  {Dieulesaint}},\ }\href@noop {} {\emph {\bibinfo {title} {{Elastic Waves in
  Solids II. Generation, Acousto-optic Interaction, Application}}}}\ (\bibinfo
  {publisher} {Springer-Verlag, Berlin},\ \bibinfo {year} {2000})\BibitemShut
  {NoStop}%
\bibitem [{\citenamefont {Shihab}\ \emph {et~al.}(2017)\citenamefont {Shihab},
  \citenamefont {Thevenard}, \citenamefont {Lema{\^{i}}tre},\ and\
  \citenamefont {Gourdon}}]{Shihab2017}%
  \BibitemOpen
  \bibfield  {author} {\bibinfo {author} {\bibfnamefont {S.}~\bibnamefont
  {Shihab}}, \bibinfo {author} {\bibfnamefont {L.}~\bibnamefont {Thevenard}},
  \bibinfo {author} {\bibfnamefont {A.}~\bibnamefont {Lema{\^{i}}tre}}, \ and\
  \bibinfo {author} {\bibfnamefont {C.}~\bibnamefont {Gourdon}},\ }\href
  {\doibase 10.1103/PhysRevB.95.144411} {\bibfield  {journal} {\bibinfo
  {journal} {Physical Review B}\ }\textbf {\bibinfo {volume} {95}},\ \bibinfo
  {pages} {144411} (\bibinfo {year} {2017})}\BibitemShut {NoStop}%
\bibitem [{\citenamefont {Foerster}\ \emph {et~al.}(2017)\citenamefont
  {Foerster}, \citenamefont {Maci{\`{a}}}, \citenamefont {Statuto},
  \citenamefont {Finizio}, \citenamefont {Hern{\'{a}}ndez-M{\'{i}}nguez},
  \citenamefont {Lend{\'{i}}nez}, \citenamefont {Santos}, \citenamefont
  {Fontcuberta}, \citenamefont {Hern{\`{a}}ndez}, \citenamefont {Kl{\"{a}}ui},\
  and\ \citenamefont {Aballe}}]{Foerster2017}%
  \BibitemOpen
  \bibfield  {author} {\bibinfo {author} {\bibfnamefont {M.}~\bibnamefont
  {Foerster}}, \bibinfo {author} {\bibfnamefont {F.}~\bibnamefont
  {Maci{\`{a}}}}, \bibinfo {author} {\bibfnamefont {N.}~\bibnamefont
  {Statuto}}, \bibinfo {author} {\bibfnamefont {S.}~\bibnamefont {Finizio}},
  \bibinfo {author} {\bibfnamefont {A.}~\bibnamefont
  {Hern{\'{a}}ndez-M{\'{i}}nguez}}, \bibinfo {author} {\bibfnamefont
  {S.}~\bibnamefont {Lend{\'{i}}nez}}, \bibinfo {author} {\bibfnamefont
  {P.~V.}\ \bibnamefont {Santos}}, \bibinfo {author} {\bibfnamefont
  {J.}~\bibnamefont {Fontcuberta}}, \bibinfo {author} {\bibfnamefont {J.~M.}\
  \bibnamefont {Hern{\`{a}}ndez}}, \bibinfo {author} {\bibfnamefont
  {M.}~\bibnamefont {Kl{\"{a}}ui}}, \ and\ \bibinfo {author} {\bibfnamefont
  {L.}~\bibnamefont {Aballe}},\ }\href {\doibase 10.1038/s41467-017-00456-0}
  {\bibfield  {journal} {\bibinfo  {journal} {Nature Communications}\ }\textbf
  {\bibinfo {volume} {8}},\ \bibinfo {pages} {407} (\bibinfo {year}
  {2017})}\BibitemShut {NoStop}%
\bibitem [{\citenamefont {Sch{\"{u}}lein}\ \emph {et~al.}(2015)\citenamefont
  {Sch{\"{u}}lein}, \citenamefont {Zallo}, \citenamefont {Atkinson},
  \citenamefont {Schmidt}, \citenamefont {Trotta}, \citenamefont {Rastelli},
  \citenamefont {Wixforth},\ and\ \citenamefont {Krenner}}]{Schulein2015}%
  \BibitemOpen
  \bibfield  {author} {\bibinfo {author} {\bibfnamefont {F.~J.}\ \bibnamefont
  {Sch{\"{u}}lein}}, \bibinfo {author} {\bibfnamefont {E.}~\bibnamefont
  {Zallo}}, \bibinfo {author} {\bibfnamefont {P.}~\bibnamefont {Atkinson}},
  \bibinfo {author} {\bibfnamefont {O.~G.}\ \bibnamefont {Schmidt}}, \bibinfo
  {author} {\bibfnamefont {R.}~\bibnamefont {Trotta}}, \bibinfo {author}
  {\bibfnamefont {A.}~\bibnamefont {Rastelli}}, \bibinfo {author}
  {\bibfnamefont {A.}~\bibnamefont {Wixforth}}, \ and\ \bibinfo {author}
  {\bibfnamefont {H.~J.}\ \bibnamefont {Krenner}},\ }\href {\doibase
  10.1038/nnano.2015.72} {\bibfield  {journal} {\bibinfo  {journal} {Nat.
  Nanotechnol.}\ }\textbf {\bibinfo {volume} {10}},\ \bibinfo {pages} {512}
  (\bibinfo {year} {2015})}\BibitemShut {NoStop}%
\bibitem [{\citenamefont {Holl{\"{a}}nder}\ \emph {et~al.}(2017)\citenamefont
  {Holl{\"{a}}nder}, \citenamefont {M{\"{u}}ller}, \citenamefont {Lohmann},
  \citenamefont {Mozooni},\ and\ \citenamefont {McCord}}]{Hollander2017}%
  \BibitemOpen
  \bibfield  {author} {\bibinfo {author} {\bibfnamefont {R.~B.}\ \bibnamefont
  {Holl{\"{a}}nder}}, \bibinfo {author} {\bibfnamefont {C.}~\bibnamefont
  {M{\"{u}}ller}}, \bibinfo {author} {\bibfnamefont {M.}~\bibnamefont
  {Lohmann}}, \bibinfo {author} {\bibfnamefont {B.}~\bibnamefont {Mozooni}}, \
  and\ \bibinfo {author} {\bibfnamefont {J.}~\bibnamefont {McCord}},\ }\href
  {\doibase 10.1016/j.jmmm.2017.01.091} {\bibfield  {journal} {\bibinfo
  {journal} {J. Magn. Magn. Mater.}\ }\textbf {\bibinfo {volume} {432}},\
  \bibinfo {pages} {283} (\bibinfo {year} {2017})}\BibitemShut {NoStop}%
\bibitem [{\citenamefont {Wei{\ss}}\ \emph {et~al.}(2018)\citenamefont
  {Wei{\ss}}, \citenamefont {H{\"{o}}rner}, \citenamefont {Zallo},
  \citenamefont {Atkinson}, \citenamefont {Rastelli}, \citenamefont {Schmidt},
  \citenamefont {Wixforth},\ and\ \citenamefont {Krenner}}]{Weiss2018}%
  \BibitemOpen
  \bibfield  {author} {\bibinfo {author} {\bibfnamefont {M.}~\bibnamefont
  {Wei{\ss}}}, \bibinfo {author} {\bibfnamefont {A.~L.}\ \bibnamefont
  {H{\"{o}}rner}}, \bibinfo {author} {\bibfnamefont {E.}~\bibnamefont {Zallo}},
  \bibinfo {author} {\bibfnamefont {P.}~\bibnamefont {Atkinson}}, \bibinfo
  {author} {\bibfnamefont {A.}~\bibnamefont {Rastelli}}, \bibinfo {author}
  {\bibfnamefont {O.~G.}\ \bibnamefont {Schmidt}}, \bibinfo {author}
  {\bibfnamefont {A.}~\bibnamefont {Wixforth}}, \ and\ \bibinfo {author}
  {\bibfnamefont {H.~J.}\ \bibnamefont {Krenner}},\ }\href {\doibase
  10.1103/PhysRevApplied.9.014004} {\bibfield  {journal} {\bibinfo  {journal}
  {Phys. Rev. Appl.}\ }\textbf {\bibinfo {volume} {9}},\ \bibinfo {pages}
  {014004} (\bibinfo {year} {2018})}\BibitemShut {NoStop}%
\bibitem [{\citenamefont {Maichen}(2006)}]{maichen2006}%
  \BibitemOpen
  \bibfield  {author} {\bibinfo {author} {\bibfnamefont {W.}~\bibnamefont
  {Maichen}},\ }\href@noop {} {\emph {\bibinfo {title} {{Digital timing
  measurements: from scopes and probes to timing and jitter}}}}\ (\bibinfo
  {publisher} {Springer Science {\&} Business Media},\ \bibinfo {year}
  {2006})\BibitemShut {NoStop}%
\bibitem [{\citenamefont {N{\v{e}}mec}\ \emph {et~al.}(2012)\citenamefont
  {N{\v{e}}mec}, \citenamefont {Rozkotov{\'{a}}}, \citenamefont
  {Tesa\v{r}ov{\'{a}}}, \citenamefont {Troj{\'{a}}nek}, \citenamefont {{De
  Ranieri}}, \citenamefont {Olejn{\'{\i}}k}, \citenamefont {Zemen},
  \citenamefont {Nov{\'{a}}k}, \citenamefont {Cukr}, \citenamefont
  {Mal{\'{y}}},\ and\ \citenamefont {Jungwirth}}]{Nemec2012}%
  \BibitemOpen
  \bibfield  {author} {\bibinfo {author} {\bibfnamefont {P.}~\bibnamefont
  {N{\v{e}}mec}}, \bibinfo {author} {\bibfnamefont {E.}~\bibnamefont
  {Rozkotov{\'{a}}}}, \bibinfo {author} {\bibfnamefont {N.}~\bibnamefont
  {Tesa\v{r}ov{\'{a}}}}, \bibinfo {author} {\bibfnamefont {F.}~\bibnamefont
  {Troj{\'{a}}nek}}, \bibinfo {author} {\bibfnamefont {E.}~\bibnamefont {{De
  Ranieri}}}, \bibinfo {author} {\bibfnamefont {K.}~\bibnamefont
  {Olejn{\'{\i}}k}}, \bibinfo {author} {\bibfnamefont {J.}~\bibnamefont
  {Zemen}}, \bibinfo {author} {\bibfnamefont {V.}~\bibnamefont {Nov{\'{a}}k}},
  \bibinfo {author} {\bibfnamefont {M.}~\bibnamefont {Cukr}}, \bibinfo {author}
  {\bibfnamefont {P.}~\bibnamefont {Mal{\'{y}}}}, \ and\ \bibinfo {author}
  {\bibfnamefont {T.}~\bibnamefont {Jungwirth}},\ }\href {\doibase
  10.1038/nphys2279} {\bibfield  {journal} {\bibinfo  {journal} {Nat. Phys.}\
  }\textbf {\bibinfo {volume} {8}},\ \bibinfo {pages} {411} (\bibinfo {year}
  {2012})}\BibitemShut {NoStop}%
\bibitem [{\citenamefont {Royer}\ and\ \citenamefont
  {Dieulesaint}(2000{\natexlab{b}})}]{Royer2000a}%
  \BibitemOpen
  \bibfield  {author} {\bibinfo {author} {\bibfnamefont {D.}~\bibnamefont
  {Royer}}\ and\ \bibinfo {author} {\bibfnamefont {E.}~\bibnamefont
  {Dieulesaint}},\ }\href
  {http://scholar.google.com/scholar?hl=en{\&}btnG=Search{\&}q=intitle:Elastic+Waves+in+Solids+I:+Free+and+Guided+Propagation{\#}1}
  {\emph {\bibinfo {title} {{Elastic Waves in Solid I: Free and Guided
  Propagation}}}}\ (\bibinfo  {publisher} {Springer-Verlag, Berlin
  Heidelberg},\ \bibinfo {year} {2000})\BibitemShut {NoStop}%
\bibitem [{\citenamefont {Saito}\ \emph {et~al.}(2010)\citenamefont {Saito},
  \citenamefont {Matsuda}, \citenamefont {Tomoda},\ and\ \citenamefont
  {Wright}}]{Saito2010}%
  \BibitemOpen
  \bibfield  {author} {\bibinfo {author} {\bibfnamefont {T.}~\bibnamefont
  {Saito}}, \bibinfo {author} {\bibfnamefont {O.}~\bibnamefont {Matsuda}},
  \bibinfo {author} {\bibfnamefont {M.}~\bibnamefont {Tomoda}}, \ and\ \bibinfo
  {author} {\bibfnamefont {O.~B.}\ \bibnamefont {Wright}},\ }\href {\doibase
  10.1364/JOSAB.27.002632} {\bibfield  {journal} {\bibinfo  {journal} {J. Opt.
  Soc. Am. B}\ }\textbf {\bibinfo {volume} {27}},\ \bibinfo {pages} {2632}
  (\bibinfo {year} {2010})}\BibitemShut {NoStop}%
\bibitem [{\citenamefont {Santos}(1999)}]{Santos1999}%
  \BibitemOpen
  \bibfield  {author} {\bibinfo {author} {\bibfnamefont {P.~V.}\ \bibnamefont
  {Santos}},\ }\href {\doibase 10.1063/1.123241} {\bibfield  {journal}
  {\bibinfo  {journal} {Appl. Phys. Lett.}\ }\textbf {\bibinfo {volume} {74}},\
  \bibinfo {pages} {4002} (\bibinfo {year} {1999})}\BibitemShut {NoStop}%
\bibitem [{Note1()}]{Note1}%
  \BibitemOpen
  \bibinfo {note} {Because we are only considering the uniform FMR mode, it is
  unecessary to take into account the absorption and optical phase shift of the
  light in the layer, as in Ref. {[}\protect \rev@citealpnum
  {Shihab2017}{]}}\BibitemShut {NoStop}%
\end{thebibliography}%
\end{document}